\documentstyle[mprocl]{article}

\begin{document}

\title{On a Theory for Nonminimal Gravitational-Electromagnetic 
Coupling Consistent with Observational Data}

\author{R. Opher} 
\address{Instituto de Astronomia e Geof\'\i sica, Universidade 
de S\~ao Paulo, \\ Av. Miguel Stefano, 4.200, CEP 04301-904, 
S\~ao Paulo, SP, Brazil}

\author{U. F. Wichoski}  
\address{Department of Physics, Brown University, \\ 
Providence, RI 02912, USA}
\maketitle 

\abstracts{
The idea that the seed primordial magnetic fields can be explained by 
the nonminimal coupling between gravitational and electromagnetic 
fields is discussed. The predicted values of the magnetic field of 
the spiral galaxies are in agreement with the observations.}

\section{Introduction}
The principle of minimal coupling (PMC) \cite{ber}, 
incorporated into Einstein's 
theory, states that the matter fields do not couple directly to curvature.
In spite of its simplicity, this principle may not lead to the correct 
theory of the interaction between gravitation and matter fields. 
Furthermore, if derivatives, higher than the first, occur in a special 
relativistic Lagrangian, the PMC does not lead to a unique Lagrangian 
in the presence of the gravitational field. Consequently, we examine
a more general approach, such as that offered by the nonminimal 
coupling (NMC) between gravitation and matter fields, 
which does not necessarily violate 
the equivalence principle \cite{acc}.

Nonminimal gravitational-electromagnetic coupling, as embodied in the 
S-B conjecture \cite{sch,bc}, 
indicates the empirical relation between the angular 
momentum ${\bf L}$ 
and the magnetic dipole moment ${\bf m}$ 
\begin{equation}
{\bf m} \cong \beta {\left[ \frac{\sqrt{G}}{2 c}\right]} {\bf L} \, , 
\label{e1}
\end{equation}
where $G$ is the Newtonian constant of gravitation, $c$ is the 
speed of light and $\beta$ is a constant.

Although, we do not yet have a satisfactory gravitational theory which 
encompasses the S-B conjecture, there is an increasing amount of 
evidence in its support. 
When we consider the magnetic field of astronomical objects, 
we note that the values calculated from Eq.(\ref{e1}) 
are in agreement with the observed values. 
Sirag~\cite{sir} compared the predictions of 
Eq.(\ref{e1}) with the observed values of the ratio 
of the magnetic moment to the angular momentum for the Earth, 
Sun, the star 78 Vir, the Moon, Mercury, Venus, Jupiter, Saturn, and 
the neutron star Her X-1. The minimum values for $\beta$ for these 
objects were: 0.12, 0.02, 0.02, 0.11, 0.37, 0.04, 0.03, 0.03, and 0.07, 
respectively. Excluding the star 78 Vir, the maximum value for $\beta$ 
was 0.77 for the planet Mercury. 
Woodward \cite{woo} examined the S-B 
conjecture in the context of pulsar gyromagnetic ratios, for 
short-period pulsars. He found that: 1) $\beta$ is not 
the same for all pulsars; 2) young pulsars evolve with their 
individual value of $\beta$, constant for a discernible period of 
time; and 3) $\beta$ lies in the range 0.001 to 0.01. 

\section{Origin of the Primordial Magnetic Fields in the Universe 
due to NMC}

The origin of magnetic fields in the Universe is one of the major 
problems in astrophysics. Magnetic fields of $\sim 10^{-6}\,\mbox{G}$ 
are observed in galaxies such as our own. Many theories have 
been proposed over the last half century for the source of the 
magnetic field. Essentially all of them have been shown 
to be unsatisfactory \cite{fie}. 

We suggest \cite{oph} that the observed 
$10^{-6}\;\mbox{G}$ magnetic field in the spiral galaxies could 
originate from the angular momentum of the protogalaxies through 
nonminimal gravitational-electromagnetic coupling as manifested 
by the S-B conjecture. 

Basically, we consider that the angular momentum of galaxies 
was acquired during the 
protogalaxy stage through the tidal torques by neighboring 
protogalaxies~\cite{pee}. 
We studied a protogalaxy with total mass 
$M \sim 10^{13} M_{\odot}$, 
corresponding to a spiral galaxy possessing a halo of dark matter 
$\sim 10$ times the mass of the luminous matter 
$M_{\mbox{L}} \sim 10^{12} M_{\odot}$. 

The angular momentum of the protogalaxy increased 
until the protogalaxies became sufficiently far apart so that 
the protogalaxy decoupled 
from the other protogalaxies, preserving its angular momentum 
${\bf L}$, acquired from the tidal interaction with the other 
protogalaxies. Up to the time of decoupling, the mean 
density of the protogalaxy was roughly that of the Universe,
${\rho}(z) = (1 + z)^{3} \rho_{0}$, 
where $\rho_{0}$ is the present matter density of the Universe 
($\rho_{0} \sim 1.057 \times 10^{-29}\;\mbox{g}\,\mbox{cm}^{-3}$ 
with $H_{0} = 75\;\mbox{km}\,\mbox{s}^{-1}\,\mbox{Mpc}^{-1}$). 

The radius of the protogalaxy ${R}(z)$ is then 
${R}(z) = [(3/4 \pi)\, M\,  {\rho}(z)^{-1}]^{1/3}$ 
and, accordingly, the angular momentum ${L}(z)$ is 
\begin{equation}
{L}(z) \cong \frac{2}{5} \lambda_{\mbox{med}} \left[ G M^{3} 
{R}(z)\right]^{1/2}\, ,
\label{e2}
\end{equation}
where $\lambda_{\mbox{med}}$ is a parameter obtained from numerical 
simulations \cite{pad}.
The magnetic field in the vicinity of the protogalaxy is obtained 
from the relation 
${{\bf B}_{\mbox{NMC}}}(z) \cong {\bf m}(z) {R}(z)^{-3}$. 
Since the magnetic field of the magnetic dipole is frozen 
into the plasma of the galaxy, which collapses to a present radius 
$R_{\mbox{L}} \simeq 10\;\mbox{kpc} \simeq 3.1 \times 10^{22}\;\mbox{cm}$, 
the present magnetic field is then obtained at the radius 
$R_{\mbox{L}}$ from \linebreak 
${\bf B}_{0}(R_{\mbox{L}}, z_{d}) \cong 
{{\bf B}_{\mbox{NMC}}}(z_{d}) [{R}(z_{d}) / R_{\mbox{L}}]^{3}$.
There is an additional amplification due to the 
differential rotation which ranges from 10 to 100. 

Hence, for a decoupling redshift $z_{d} \leq 10$, we obtain the 
present magnetic field
\begin{equation}
{{B}_{0}}(R_{\mbox{L}}, z_{d}) \sim 
10^{-6} \mbox{---} 10^{-5}\;\mbox{G}\, ,
\label{e3}
\end{equation}
where we use $\beta = 0.1$ in Eq.(\ref{e1}).

For $\beta = 0.01$ with an 
amplification due to differential rotation of 100, we obtain the same 
values of ${B_{0}}(R_{\mbox{L}}, z_{d})$ as given 
in relation (\ref{e3}). In both cases, the results are consistent 
with the ranges obtained by Sirag and Woodward. 

One might imagine that the process described here could explain 
the dark matter problem in galaxies as well. The idea is that  
matter spinning around the galaxy at a distance of approximately 
50 kpc would be subject to an extra attraction force due to the 
magnetic field generated by its own rotation and to the rotation of 
the galaxy; both via Eq.(\ref{e1}). However, we found that 
the intensity of the extra force is of order $10^{-40} \mbox{dyne}$. 
This intensity is too small to account for the rotation curves 
without considering dark matter.

\section{Conclusions}

The increasing amount of evidence in favor of the S-B conjecture in 
the astrophysical domain indicates that perhaps gravitation is not 
minimally coupled to electromagnetic fields. 
In general, the NMC is believed to be possible only in regions of 
strong gravitational fields. The evidence, however, points in the 
direction that NMC may exist even though strong gravitational 
fields are not involved.

\section*{Acknowledgments}

R.O. would like to acknowledge the partial support of 
the Brazilian agency CNPq; 
U.F.W. the partial support of the US 
Department of Energy under grant DE-F602-91ER40688, Task A.


\section*{References}

\begin{thebibliography}{99}

\bibitem{ber} P. G. Bergmann, Intern. J. Theor. Phys. 
{\bf 1}, 25 (1968); H. F. M. Goenner, Found. Phys. {\bf 14}, 865 
(1984); M. Novello and L. A. R. Oliveira, Rev. Bras. Fis. 
{\bf 17}, 432 (1987).
\bibitem{sch} A. Schuster, Proc. R. Inst. {\bf 13}, 273 (1890); 
Proc. Phys. Soc. London {\bf 24}, 121 (1912).
\bibitem{bc} P. M. S. Blackett, Nature {\bf 159}, 658 (1947); 
P. M. S. Blackett, Philos Trans. R. Soc. London A 
{\bf 245}, 309 (1952).
\bibitem{acc} A. J. Accioly and U. F. Wichoski, Class. Quantum Grav. 
{\bf 7},  L139 (1990).

\bibitem{sir} S.-P. Sirag, Nature {\bf 278}, 535 (1979).
\bibitem{woo} J. F. Woodward, Found. Phys. {\bf 19}, 1345 (1989).
\bibitem{fie} G. Field, {\it in} Proceedings of the International 
Congress of Plasma Physics, Foz de Igua\c{c}u, 1994 (AIP, New
York, 1995); R. Opher, {\it ibid};
J. H. Piddington, Aust. J. Phys. {\bf 23}, 731 (1970); 
R. M. Kulsrud, {\it in} IAU Symp. 140, Galactic and 
Intergalactic Magnetic Fields, ed. R. Beck {\it et al.} (Kluwer, 
Dordrecht, 1990), p. 527; 
J. H. Piddington, {\it Cosmic Electrodynamics}, 
(Krieger, Malabar, 1981), 2nd ed..
\bibitem{oph} R. Opher and U. F. Wichoski, Phys. Rev. Lett. 
{\bf 78}, 787 (1997).
\bibitem{pee} P. J. E. Peebles, Astrophys. J. {\bf 155}, 393 (1969); 
S. D. M. White, Astrophys. J. {\bf 286}, 38 (1984); 
G. Efstathiou and B. J. T. Jones, Mon. Not. R. Astron. 
Soc. {\bf 186}, 133 (1979); 
J. Barnes and G. Efstathiou, Astrophys. J. 
{\bf 319}, 575 (1987).
\bibitem{pad} T. Padmanabhan, {\it Structure Formation in 
the Universe}, (Cambrigde University Press, Cambridge, 1993).
\end{thebibliography}
\end{document}